\documentclass[final,5p,times]{elsarticle}
\usepackage{graphicx,color}
\usepackage{amssymb}
\usepackage[nodots]{numcompress}

\usepackage{lineno}
\journal{Axiv}

\usepackage{amsfonts,amsbsy,graphicx,verbatim,color}
\usepackage[noend]{algorithmic} 
\usepackage{algorithm}%,caption}
\algsetup{indent=2em} 
\renewcommand{\algorithmiccomment}[1]{\hspace{2em}// #1} 
\renewcommand{\algorithmicrequire}{\textbf{Input:}}
\renewcommand{\algorithmicensure}{\textbf{Output:}}

\newtheorem{thm}{Theorem}[section]
\newtheorem{crl}{Corollary}[section]
\newtheorem{Defi}{Definition}[section]

\newcommand{\C}{{\cal C}}
\newcommand{\E}{{\cal E}}
\newcommand{\J}{{\cal J}}
\renewcommand{\L}{{\cal L}}
\renewcommand{\P}{{\cal P}}
\newcommand{\R}{\mathbb{R}}
\newcommand{\Z}{\mathbb{Z}}

\newcommand{\la}{\leftarrow} 
\newcommand{\BE}{\begin{equation}}  
\newcommand{\EE}{\end{equation}}

\begin{document}

\begin{frontmatter}
\title{A programme to determine the exact interior of any connected digital picture}
\author[label1]{A. E. Fabris\corref{cor1}}
%\cortext[cor1]{Corresponding author. Tel: +55-11-30916216.}
%\ead{aefabris@gmail.com}
%\ead[url]{https://sites.google.com/site/aefabris}
\author[label2]{V. Ramos Batista}
\address[label1]{USP - IME, r. do Mat\~ao 1010, 05508-090 S\~ao Paulo-SP, Brazil}
\address[label2]{UFABC - CMCC, r. Sta. Ad\'elia 166, Bl.B, 09210-170 St. Andr\'e-SP, Brazil}
\begin{abstract}
Region filling is one of the most important and fundamental operations in computer graphics and image processing. Many filling algorithms and their implementations are based on the Euclidean geometry, which are then translated into computational models moving carelessly from the continuous to the finite discrete space of the computer. The consequences of this approach is that most implementations fail when tested for challenging degenerate and nearly degenerate regions. We present a correct integer-only procedure that works for all connected digital pictures. It finds all possible interior points, which are then displayed and stored in a locating matrix. Namely, we present a filling {\it and} locating procedure that can be used in computer graphics and image processing applications.
\end{abstract}
\begin{keyword}
Raster Graphics\sep Region Filling\sep Point-based Modelling\sep Discrete Jordan Curve Theorem
\end{keyword}
\end{frontmatter}

%\linenumbers

\section{Introduction}
\label{intro}

Region filling is a crucial operation in computer graphics and image processing. There are two basic approaches to region filling on raster systems. One of them is to start from a given interior position called {\it seed}, and paint outward from this point until we meet the specified boundary conditions. The other is to determine the overlap intervals of scan lines that cross the region. The seed-fill methods are mostly applied in interactive painting systems. Alternatively, the scan-conversion approach is typically used in hardware implementations and in general graphic packages to fill polygons, circumferences, ellipses and spline curves. These two approaches will be resumed in Section~\ref{seedscan}, but now we focus on their limitations: the non-robustness of scan conversion, and the high computational cost of seed fill.

As explained in Section~\ref{seedscan}, neither seed fill nor scan conversion can really provide perfect filling at the lowest computational cost. This is yet desirable in many circumstances. We list three of them.

\begin{figure}[ht!]
\centerline{
\includegraphics[scale=0.41]{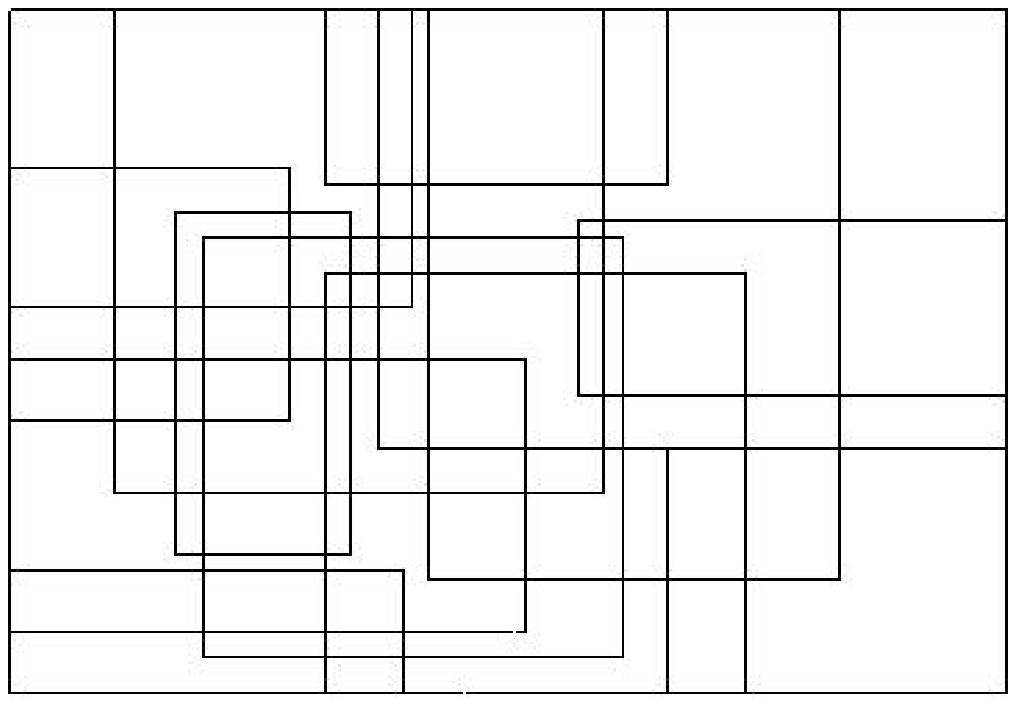}
\includegraphics[scale=0.30]{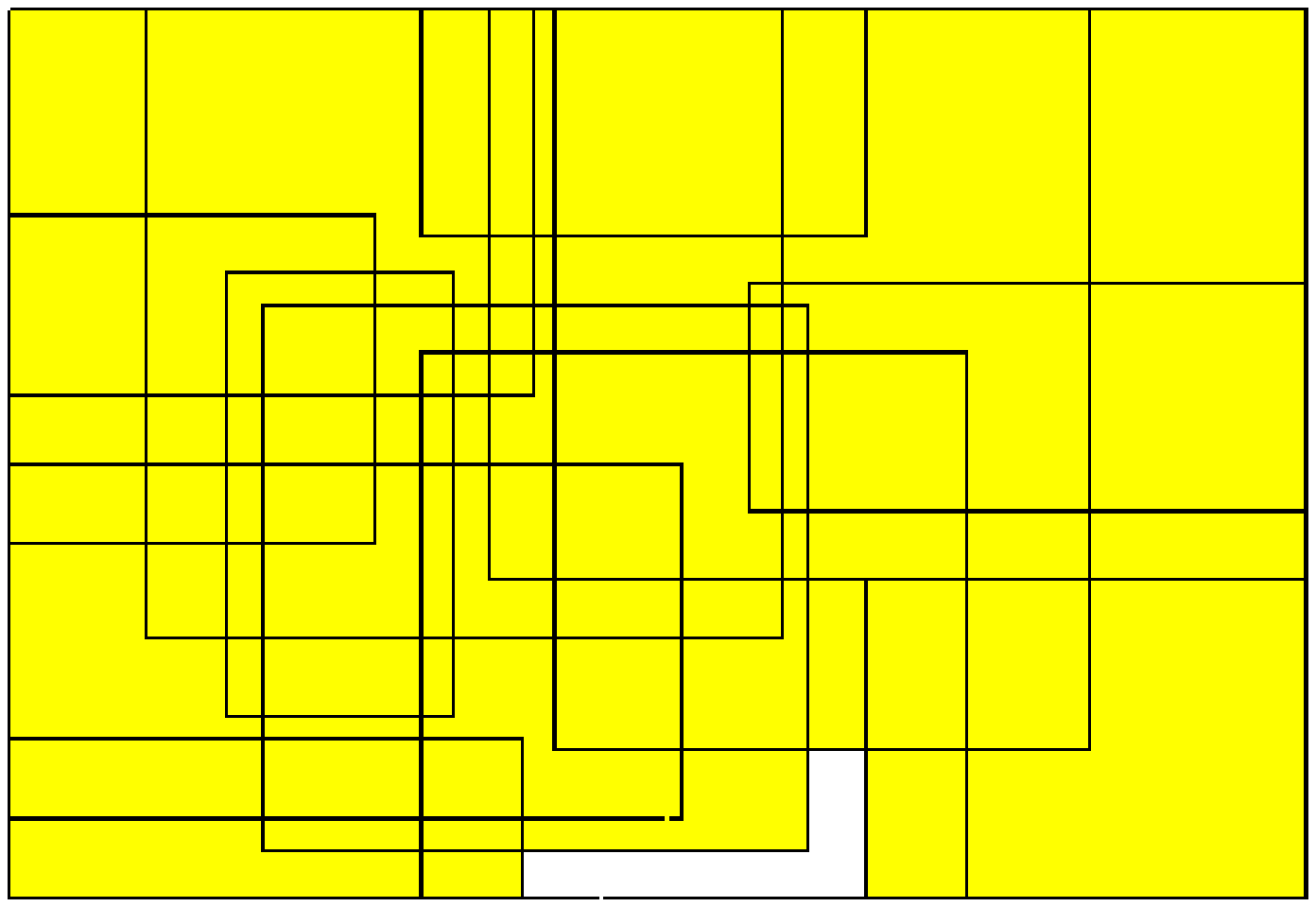}}
\caption{Detecting an imperceptible flaw in a mould of integrated circuit.}
\label{circuit}
\end{figure}

The first is represented in Figure~\ref{circuit}. It schematises the mould of an integrated circuit. In general, thousands of models are manufactured from the mould. A single flaw in it will produce useless lots, and therefore a great waste of material and time. Until nowadays, this kind of flaw detection still depends on handwork. 

In fact, Figure~\ref{circuit} right is an output of our algorithms. They aim at filling regions perfectly and quickly. As an application, this helps detect imperceptible flaws. For the time being, however, they can detect only {\it external} flaws, but are adaptable for the whole job.

Our second example is of theoretical interest. Of course, there is already a valuable and extensive literature about region filling and point location. In fact, both problems are intimately related. Now, according to \cite{Sch}, none of them is free of minor failures, and the best ones are computationally expensive. Therefore, if there are algorithms that prove to be quick and faultless, they are really worth studying.

The third and most striking example is the technology of 3D-printing. This paper deals with a two-dimensional approach. But our algorithms were elaborated with ideas that {\it are} extendable to higher dimensions. In fact, we are already researching this problem for future works. The basic idea is to slice a 3-dimensional object into several 2-dimensional ones, which are then treated individually. In this case, least computational time becomes highly important. But also perfect filling and point location. The problem of printing plastic cups, for instance, is equivalent to what Figure~\ref{circuit} illustrates.

%To the best of our knowledge, all drawing softwares are susceptible to failures. Figure~\ref{panda} illustrates this problem. 
%Of course, classical algorithms for region filling and point location {\it are} of great value. In many cases we can content ourselves with flaws.

\begin{figure}[ht!]
\centerline{
\includegraphics[scale=0.35]{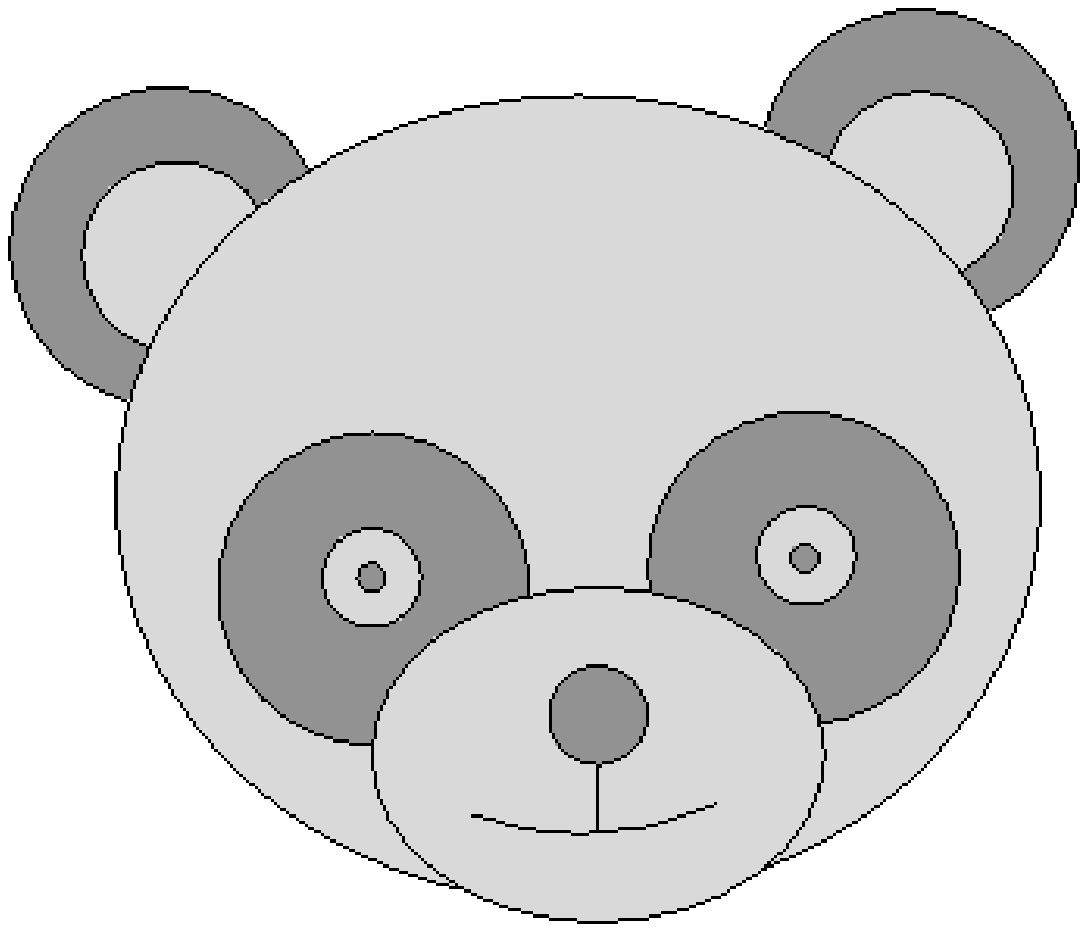}\hfill  
\includegraphics[scale=1.90]{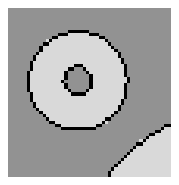}}
\caption{(a) A panda's drawing; (b) detail of area around one eye showing individual pixels.}
\label{panda}
\end{figure}

Of course, when performing region filling or point location we can content ourselves with flaws as we see in Figure~\ref{panda} drawn in Xfig and saved as a PostScript file.
If not, seed fill reduces them to a minimum but usually with a high computational cost.

Our research was motivated by the fact that scan-conversion algorithms  are not always reliable. In the case of point location algorithms, they fail even for non-challenging regions, as demonstrated in~\cite{Sch,Ket}.
Moreover, in~\cite{Sch} Schirra presents a complete comparison among all classical point-location algorithms.

In this paper, the difficulties in discretisation are acknowledged and tackled by casting the problem to be solved as a discrete integer problem from the very outset. 
We believe that region filling and point location algorithms cannot be always reliable, unless these problems are set up and solved in a discrete space. Hence, we {\it never} have to accommodate any inconvenience of numerical accuracy that follows from floating point discretisation, geometric approximation, or less rigorous approach to final discretisation in the rendering process. We give formal definitions of the involved discrete topology, together with a mathematical proof of robustness.

However, we {\it do not} start from mathematical models. Our input is solely a matrix with entries 0 and 1, which we interpret as a picture of black pixels surrounded by white pixels. No equations are given beforehand. Then we {\it construct} a mathematical model that retrieves our input, performs the region filling and generates a filling matrix. This matrix locates any point as belonging to the interior, exterior or to the picture itself. 

Our work is organised as follows. Section~\ref{seedscan} gives some background on the two classical region filling algorithms. Section~\ref{mathstr} explains some basic terms and definitions. They are used in Section~\ref{reductions}, where we discuss our main programme and its algorithm. The main programme is endowed with an optional improvement, which is explained in Section~\ref{lego}. This improvement will make the main programme correct, and Section~\ref{mainthm} presents a theoretical proof of this fact. Details about downloading and running our programme are explained in Section~\ref{ap1}, whereas Section~\ref{concl} is devoted to our conclusions. Finally, \ref{ap2} describes some theoretical results that can be achieved through the source codes.

\section{Seed fill and scan conversion}
\label{seedscan}

Seed-fill algorithms can be subdivided into four steps \cite{FisBar}: a {\it start procedure} to choose an interior starting point, namely the seed; a {\it propagation method} to determine the next considered point; an {\it inside test} to check whether a point is in the region and if it should be marked; a {\it set procedure} to change the colour of a given picture element or {\it pixel}. The {\it start procedure} can be done automatically or user-guided. For complex objects, multiple seeds may be required and their automatic detection can be very difficult. This is one of the main reasons why this family of algorithms is more suitable for interactive region filling~\cite{CodNev}. For the {\it propagation method}, the na{\" i}ve implementation of the algorithm searches for connected pixels recursively, and can be based either on pixel- or line-adjacency. In the former, the filling procedure is to move from a current pixel to all immediate neighbours in a certain order. In the latter, the interior of a region is regarded as a set of adjacent horizontal line segments. Usually, seed-fill algorithms need working memory in addition to the frame buffer, and are slower than scan-conversion algorithms. The amount of working memory cannot be constant, for it increases with both size and complexity of the region. Consequently, it cannot be stored in advance. Compared to seed-fill algorithms, scan-conversion algorithms are generally faster and require less or no additional working memory. In order to overcome the difficulties of seed-fill algorithms, different strategies like~\cite{TsaChu,LejHao} try to re-model the region and so get a fast image manipulation, processing and displaying. 

Regarding scan conversion, well-known related to algorithms require computationally expensive presorting and marking phases before they compute the actual intervals of points contained in the region. This holds even for simple concave polygons. For curved boundaries, the marking phase requires careful attention to both geometric and numerical details to provide robust algorithms. Conventional filling of curved regions is based on algorithms for scan conversion of polygons, in which the end points of spans are incrementally updated, and then the intermediate pixels are filled. At some stage the intersection formula, derived in the continuous plane and usually computed using real arithmetic, has to be mapped to the discrete plane. This mapping needs an implicit or explicit epsilon-test that may cause incorrect results. Many scan-conversion algorithms fail to handle complex objects correctly, while in principle seed filling is more robust~\cite{Sch,CodNev}.
 
Levoy and Whitted suggest points rather than triangle meshes as display primitives~\cite{LevWhi,RusLev,WeyGro}. Similarly, we propose points as low-level modelling primitives, and so get rid of the well-known discretisation problems of the polygon and of all general curved boundaries too. Points are building bricks of geometry discretisation, and they can also provide a common format at the lowest level for both constructed and scanned geometric data. Namely, for both computer graphics and image processing. 

\section{The mathematical structure}
\label{mathstr}

\begin{figure}[ht!]
\centerline{
\includegraphics[scale=0.65]{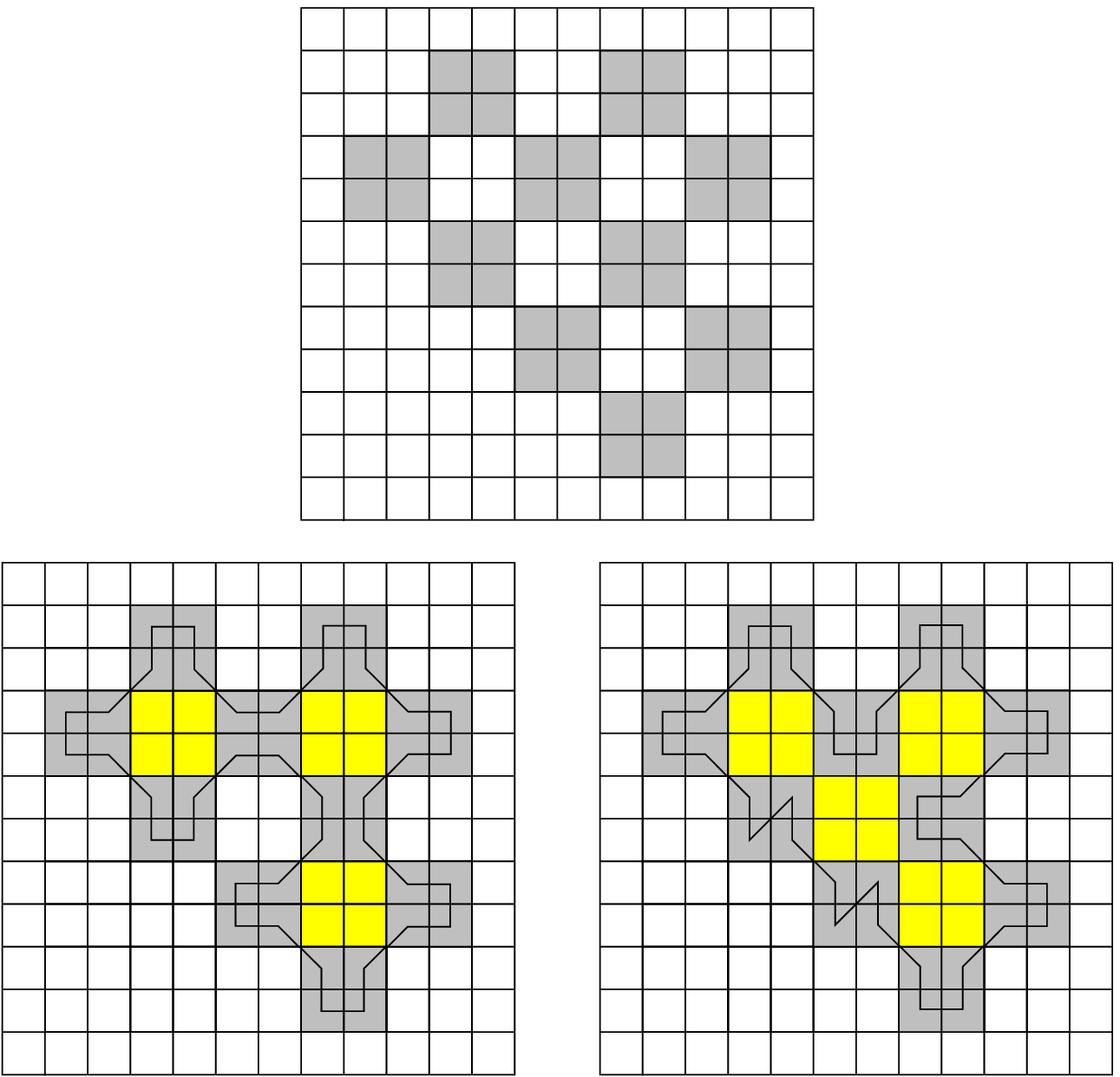}}
\caption{Ambiguity regarding interior and exterior of a digital picture (example with a $12\times12$ matrix of pixels).}
\label{f1}
\end{figure}

A dominant reason for which raster graphics algorithms fail in robustness is the absence of formal specifications. Raster systems are discrete devices, which generate image by displaying the intensity value of each pixel in a finite two-dimensional matrix of pixels. But discrete definitions are implicitly assumed to be like their respective counterparts of the continuous Euclidean space. Commonly, a formal definition of interior/exterior of digital objects is omitted. This can cause difficulties in differing between the rules of odd-parity and nonzero winding number, and mainly in comparing different algorithms. Figure \ref{f1} shows a typical ambiguity that arises when such terms are treated only implicitly.

Now we define some basic terms, and the first is our space of points:
{\Defi{\rm Given}\label{canvas}} any two positive natural numbers $N$ and $M$, a {\it canvas} in $\Z^2$ is the set $\{(p_y,p_x):1\le y\le N,1\le x\le M\}$, where $N\times M$ indicates its {\it dimension}. 
\\

In a computing context, a digital image is a two-dimensional array of values reflecting the brightness and colours of the pixels. 

{\Defi{\rm Let}\label{figure}} $S=S_{N\times M}$ be a canvas. For any $f:S\to\{0,1\}$, we say that the set $f^{-1}(1)$ is a 
{\it digital picture} 
% {\it digital curve} 
in the canvas. 
\\

In Definition \ref{figure}, our convention is to think of $0$ as white and $1$ as black. 
Therefore, the {\it digital picture} is a set of black dots over a white background. 
% we start with a fully white canvas and get our digital curve by marking black dots on it. 
%%
As remarked in the introduction, our input is a digital picture $f^{-1}(1)$, or just a matrix with entries $0$ and $1$. Our filling algorithm does not require any other mathematical expression for the function $f$.

{\Defi{\rm A}\label{dcurve}} {\it discrete curve} in a canvas $S$ is an 8-connected list $L:[0\dots n]\to S$. We say that $L$ is {\it monotone and closed} when $L(i)=L(j)\iff i=j$ mod $n$. 

{\Defi{\rm Consider}\label{cfig}} a digital picture $\P$ with its corresponding $f:S\to\{0,1\}$. If any two points $p,q\in f^{-1}(1)$ always admit a discrete curve $L:[0\dots n]\to S$ with $L(0)=p$, $L(n)=q$ and $f(L(i))=1$ $\forall i$, then we say that $\P$ is {\it connected}.

These definitions are comparable to those given in works by Rosenfeld and Kong \cite{Ros,KonRos}. 
Notice that the same digital picture can correspond to distinct discrete curves. By looking at Figure \ref{f1}, the black pixels on either case form exactly the {\it same} digital picture $\P$. However, we can {\it track} these pixels in different ways. Each image comes from a different discrete curve that represents $\P$.

\begin{figure}[ht!]
\centerline{
\includegraphics[scale=0.38]{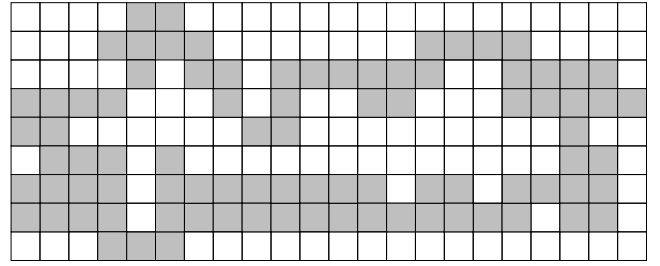}}
\caption{A digital picture with thickness varying from 1 to 3.}
\label{f2}
\end{figure}
Figure \ref{f2} helps understand the following definition:

{\Defi{\rm Given}\label{tck}} $p\in L([0\dots n])$, consider the biggest square $Q$ such that $p\in Q\subset L([0\dots n])$. Its dimension is $k\times k$ for a certain $k\ge1$. Then $k$ is the {\it thickness of the discrete curve} $L$ {\it at} $p$. 
\\

{\Defi{\rm We}\label{thin}} say that a discrete curve $L$ is {\it thin} when it has thickness 1 everywhere.
\\

In practice, Definition \ref{thin} can be replaced by a weaker condition:

{\Defi{\rm An}\label{lthin}} injective discrete curve $L:[0\dots n]\to S$ is {\it locally thin} when the following holds. There are intervals $I_1\cup\dots\cup I_k=[0\dots n]$ such that $|I_j|\ge 4$ and $L|_{I_j}$ is thin, $\forall\,j\in\{1,\dots,k\}$.
\\

For instance, the black pixels in Figure \ref{f2} cannot represent the image of an $L$ that is locally thin. We shall give more details in Section~\ref{lego}.

Our purpose is to consider {\it any} digital picture $\P$, solely restricted to Definition \ref{cfig}. For instance, take a more general example like in Figure \ref{f2}. Now the question is: How to describe our input, a digital picture $\P$ as in Figure \ref{f2}, in terms of a function $L$ in accordance with Definition \ref{dcurve}? 

This question will be answered in Section \ref{lego}. In fact, that section introduces what we call {\it the Lego curve}. Its set of pixels will not always coincide with $\P$. However, starting from the Lego curve it is then possible to obtain a function $L:[0\dots n]\to S$ as in Definition \ref{dcurve}, such that $L([0\dots n])\equiv\P$.

Notice that Definition \ref{dcurve} {\it does not} require a discrete curve to be {\it simple}. Namely, it can have {\it self-intersections} defined as follows:

{\Defi{\rm Based}\label{simple}} on Definition \ref{dcurve}, we say that $L$ has 
\\
$\bullet$ \ a {\it self-crossing} when there are two disjoint intervals $I,J\subset$ 

$[0,n]$, $|I|=|J|=2$, such that $L(I)\ne L(J)$ and all pixels 

$L(I\cup J)$ are 8-connected.
\\
$\bullet$ \ an {\it overlapping} when we have $I,J\subset[0,n]$, $\min\{|I|,|J|\}\ge$ 

$2$, such that $I\cap J$ is unitary, and either $L(I)\subset L(J)$ or 

$L(J)\subset L(I)$.
\\
If any of these two cases occur, then $L$ has a {\it self-intersection}. Otherwise, it is a {\it simple} curve.
\\

Examples in $\R^2$ can clarify Definition \ref{simple}. For instance, the so-called {\it Bernoulli's Lemniscate} is a closed curve, of which the parametric equations are $x(t)=2\cos t/(1+\sin^2t)$, $y(t)=\sin2t/(1+\sin^2t)$, with $t\in[-\pi,\pi]$. Its trace is $\infty$ (the symbol of infinity). In this case we just have a self-crossing at the origin. For the curve $(0,\sin t)$, $t\in[0,\pi]$, overlapping occurs at $t=0$ and at $t=\pi/2$. Notice that $t$ could vary in a bigger domain, but we always take the {\it smallest} for which the curve is closed. 

The reader must have noticed that we treat some technical terms as synonyms: point and pixel, discrete and digital, etc. In fact, this started to happen right at Definition \ref{canvas}, because our constructions are devoted to both theoretical and practical contexts. The reader is free to choose any of these contexts, and then think in accordance with the corresponding terms. 

\section{The main programme}
\label{reductions}

In this section we discuss the algorithm of our main programme, named {\tt loci.m} after a mnemonic to ``location of curve interior''. In Section \ref{ap1} the reader will find details about downloading and running it. The programme {\tt loci.m} is an implementation of the ``Filling Up Algorithm'' (FUA). Before discussing FUA, it is important to explain some hypotheses under which we claim that {\tt loci.m} will always give a correct answer. 

We recall the difference between two similar terms: {\it connectedness} and {\it connectivity}. Connectedness of a digital picture $\P$ is a global property implied by Definition \ref{cfig}. Connectivity is a local property that applies to each pixel with respect to the ones that surround it in a digital picture $\P$.

Namely, given any $p\in\P$, if either $p\pm(1,0)$ or $p\pm(0,1)$ is again in $\P$, then it has connectivity 4 with $p$. If any of $p+(\pm1,\pm1)$ is in $\P$, then it has connectivity D with $p$. Finally, when both 4- and D-cases apply to $p$, we say in short that surrounding pixels have connectivity 8 with $p$. 

Now we summarise the {\it three} conditions on $L$ that will make FUA correct. The third condition will be explained in the sequel:

{\bf C1}. $L$ is locally thin;

{\bf C2}. $L$ can only be tracked horizontally and vertically. In other words, $L:[0\dots n]\to S$ is such that $L(i)$ and $L(i+1)$ are 4-connected $\forall\,i<n$;

{\bf C3}. $L$ is ``spike-free''.
\ \\

Before discussing {\bf C3}, we need to introduce a concept called {\it $i-o$ pixels} (see Definition \ref{iopixels}) 
which is  analogous regarding scan-conversion algorithms to the usual marking phase required  to compute the intervals of points contained in the region.

\begin{figure}[ht!]
\centerline{
\includegraphics[scale=0.38]{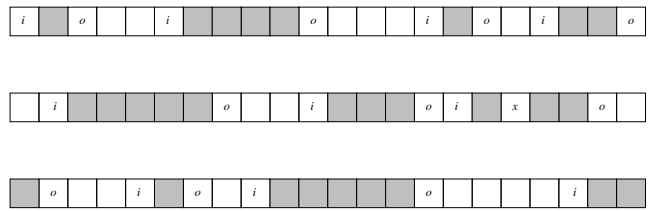}}
\caption{Some typical rows of a digital picture and their entry-exit pixels.}
\label{f3}
\end{figure}

From upside down, the scanning of a digital picture gives an $N\times M$ matrix of black and white pixels. Typical rows are represented in Figure \ref{f3}. We track each row from the left to the right. In this process, $i$-pixels indicate the entry to and $o$-pixels the exit from $\P$. When a pixel is simultaneously $i$ and $o$, we mark it with an $x$. 

In Section \ref{lego} we are going to introduce CoTRA, Connectivity and Thickness Reduction Algorithm. CoTRA produces our ``Lego curve'' $L$. 

\begin{figure}[ht!]
\centerline{
\includegraphics[scale=0.40]{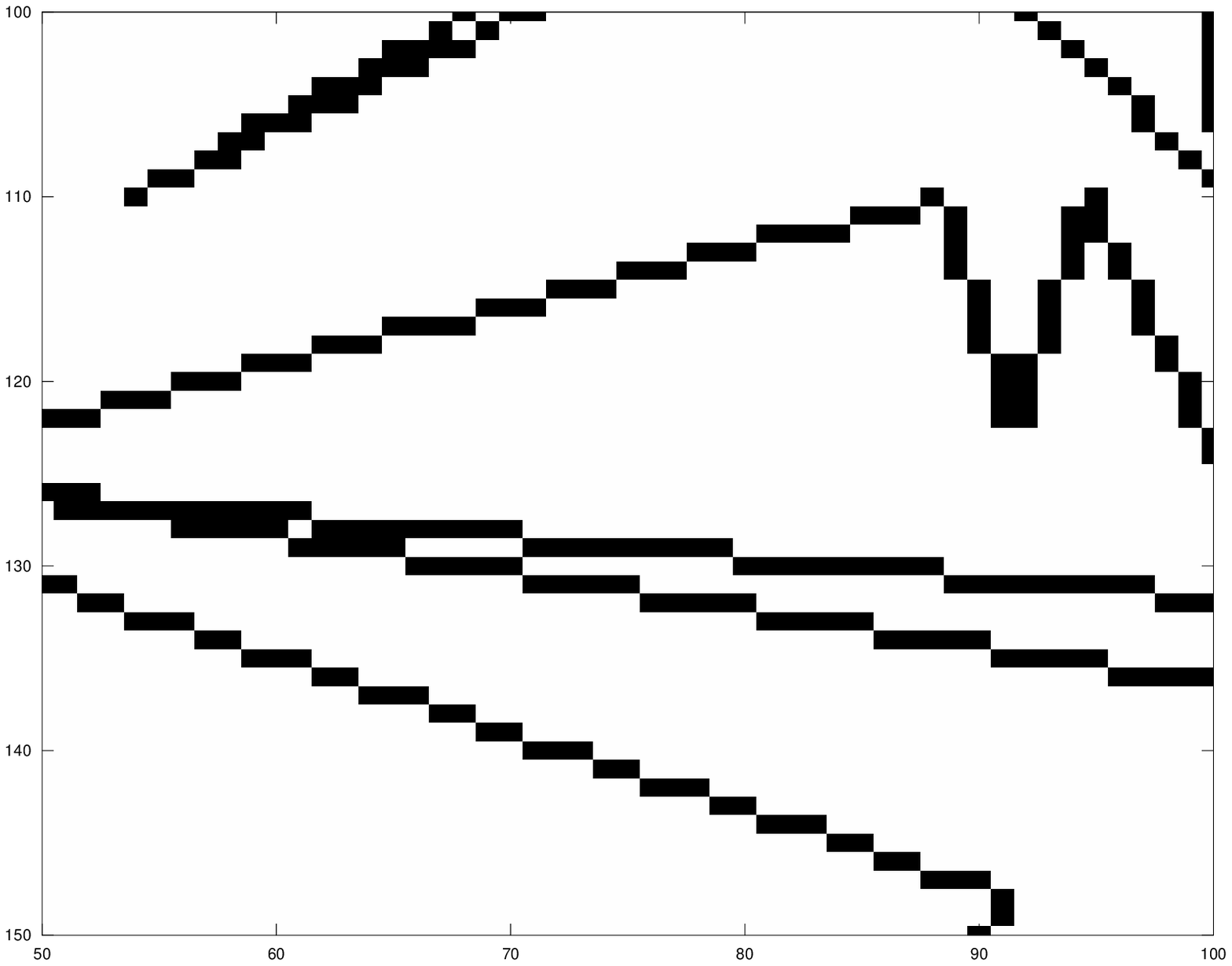}}
\end{figure}
\begin{figure}[ht!]
\centerline{
\includegraphics[scale=0.40]{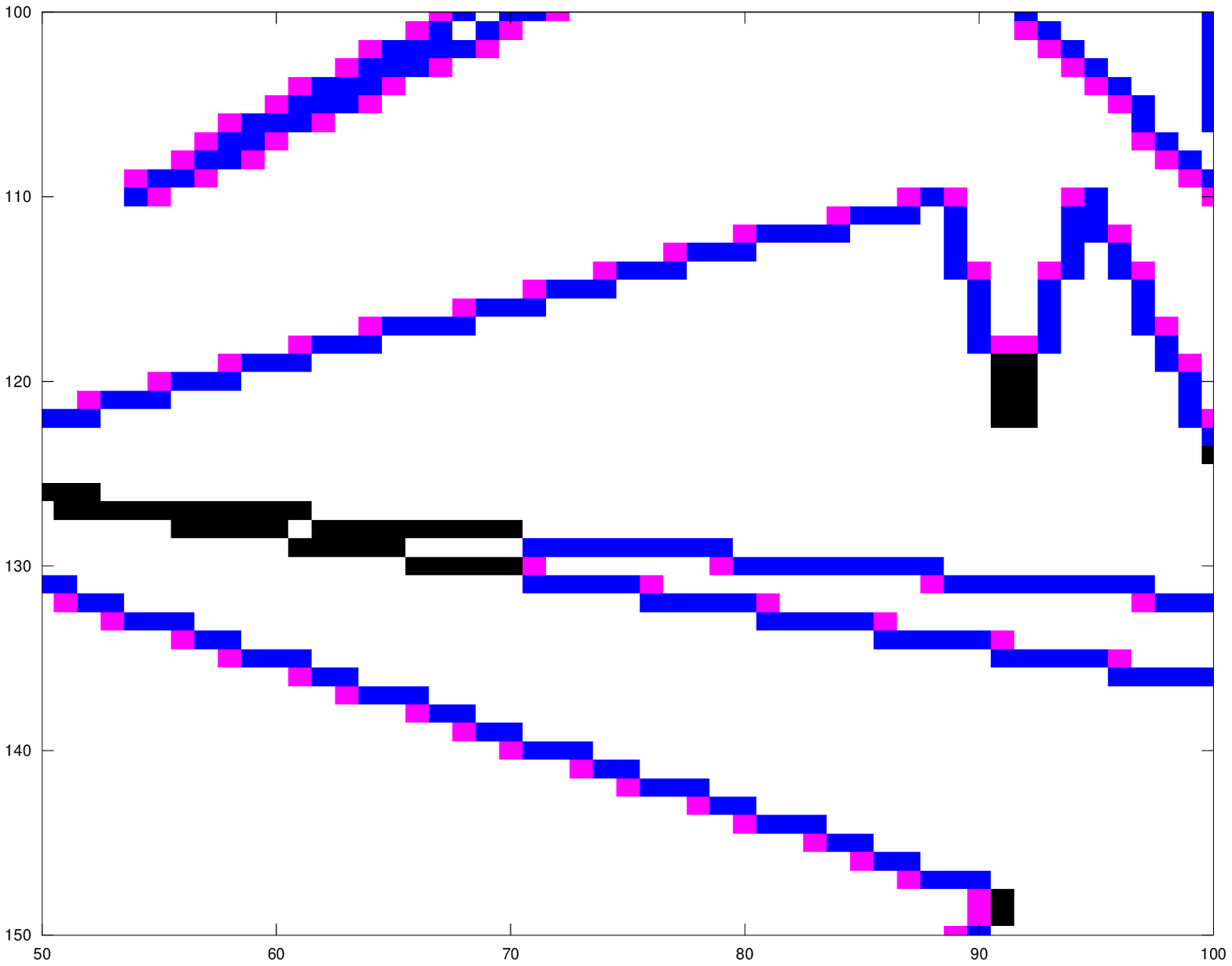}}
\caption{Input (top) and output (bottom) of the Connectivity and Thickness Reduction Algorithm (CoTRA).}
\label{f4}
\end{figure}

Figure \ref{f4} gives an example of the input (top) and the output (bottom) by CoTRA applied to a digital picture $\P$. In fact, $\P$ is a detail of Figure \ref{f8} obtained from the drawing of a non-self-intersecting polygon in Xfig. The spurious self-intersection  at the left middle occurs due to discretization errors when moving from the ideal mathematical model to the digital picture.

On top, $\P$ is given as a set of black pixels in the white canvas. At the bottom, we mark blue pixels that are originally from $\P$, and $i-o$ pixels in magenta. The blue and magenta pixels form the sought after Lego curve $L$. However, notice that not all pixels were marked by CoTRA. For example, close to $(y,x)=(130,70)$ it found a stretch interpreted as a self-intersection of $\P$. Moreover, close to $(120,90)$ and $(150,90)$ we have turning points. CoTRA cuts them by keeping local thickness 1. We assert that FUA will be correct for Lego curves $L$ (see Section \ref{mainthm} for details).

In Section \ref{lego}, we formalise ``interior'' of $\P$ by means of Definition \ref{interior}.

Now the reader may be curious about the turning points of Figure \ref{f4}. Of course, the one close to $(120,90)$ could also have thickness 1 and look like a thinner ``spike''. In technical terms, we have

{\Defi{\rm Let}\label{spike}} $L:[0\dots n]\to S$ be a monotone discrete curve with $L(i)=(L_y(i),L_x(i))\,\forall\,i$. For all $-3<k<3$, if there is $i$ such that $|L_x(i+k)-L_x(i)|<1$ and either $L_y(i)\ge L_y(i+k)$ or $L_y(i)\le L_y(i+k)$, then $L$ has a {\it vertical spike} at $i$. If there is $i$ such that $|L_y(i+k)-L_y(i)|<1$ and either $L_x(i)\ge L_x(i+k)$ or $L_x(i)\le L_x(i+k)$, then $L$ has a {\it horizontal spike} at $i$. 
\\

Handling a picture with spikes proved to needlessly increase the complexity of FUA. Indeed, we are going to obtain the interior of $\P$ as the interior of $L$ {\it minus} the set $\P$ itself. CoTRA was programmed in order to eliminate spikes and then correct FUA. 

As in any improvement, CoTRA adds a computational cost. But the users can choose: if the first output is not satisfactory, they call CoTRA for a second answer.

Now we present the basic algorithm to fill up $\P$ by marking its entry-pixels (see Algorithm~\ref{alg:FUA}). 

\begin{algorithm}
\caption{The  ``Filling Up Algorithm'' (FUA)} \label{alg:FUA}
\begin{algorithmic}[1]
\REQUIRE Digital Picture $\P=f^{-1}(1)\subset img(:,:,1)$
\ENSURE Interior of $\P$
\FOR{$i=2:$length$(img(:,1,1))-1$}
     \STATE $bps\la$ Find\_ioPixels
     \FOR{$k=1:2:$length$(bps)-1$}
        \IF{$img(i,bps(k):bps(k+1),1)$ is an extreme}
           \STATE shorten $bps$
        \ENDIF
      \ENDFOR
\ENDFOR
\FOR{$k=2:4:$length$(bps)-1$}
    	\STATE $img(i,bps(k):bps(k+1),3)=0$
\ENDFOR
\end{algorithmic}
\end{algorithm}

Notice that the ``spike-free'' condition is important in lines 4-5, where we ``jump'' the $i-o$ pixels of a local maximum (or minimum) stretch of the picture. This is because the status of being inside/outside the picture does not change at passing through such stretches.

For the rest, lines 6-7 simply mark the {\it actual} interior points of the $i$th row in yellow (R = 255, G = 255, B = 0). In the next section, we shall formalise several terms already used herein.  

\section{Definition of the Lego curve}
\label{lego}

At any row of pixels, the digital picture has white and black ones. Fix any row and consider the pixels in it. Up to scaling, each of then has an $x$-coordinate that ranges from a natural number to its successor.

{\Defi{\rm Given}\label{iopixels}} a row $R$ of black and white pixels, we say that $p\in R$ is an {\it i-pixel} when $p$ is white and $p+1$ is black. We say that $p\in R$ is an {\it o-pixel} when $p$ is white and $p-1$ is black. If $p$ is simultaneously $i$ and $o$, we call it an {\it x-pixel}.
\\

\begin{figure}[ht!]
\centerline{
\includegraphics[scale=0.33]{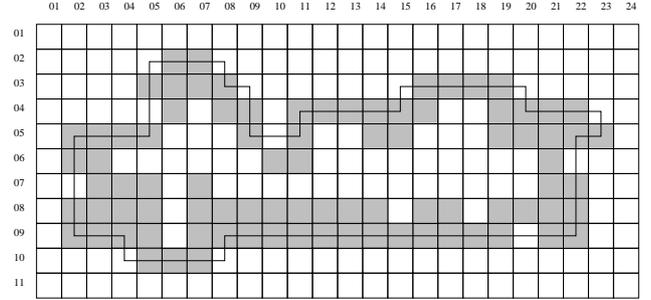}}
\caption{A digital picture and its corresponding track of the Lego curve.}
\label{f6}
\end{figure}

The letters $i$ and $o$ stand for the process of getting inside/ outside the black horizontal segments from Figure \ref{f2}, as depicted in Figure \ref{f3}. Definition \ref{iopixels} implies that the $i$-$o$ sequence is always alternated, if not empty. By adding a frame of white pixels around the picture, we guarantee the sequence to begin with $i$ and end with $o$. We have added this frame in Figure \ref{f6}.

We recall Definition~\ref{lthin}. In Figure \ref{f6}, the black pixels form a picture $\P$. Notice that $\P$ cannot be the image of a discrete curve that is locally thin. This is only possible if we remove at least 9 pixels from $\P$, namely $(7,7)$ and $(8,7-14)$.

In order to define the Lego curve, we first need to introduce two concepts as follows:

{\Defi{\rm Consider}\label{exterior}} a pixel canvas $S$ with dimension $N\times M$, and also a digital picture $\P$ contained therein. Suppose $\P$ is surrounded by the white pixel frame ${\cal F}:=\{1,N\}\times [1,M]\cup[1,N]\times\{1,M\}$. In this case, the {\it exterior} of $\P$ is the set $\E$, ${\cal F}\subset\E\subset(S\setminus\P)$, such that for any two points $p,q\in\E$ there exists a discrete curve $L:[0\dots n]\to(S\setminus\P)$ with $L(i)$ and $L(i+1)$ 4-connected $\forall\,i<n$.

{\Defi{\rm Under}\label{interior}} the hypotheses of Definition \ref{exterior}, if $\E$ is the exterior of $\P$, then its {\it interior} is $S\setminus(\P\cup\E)$.

{\Defi{\rm Let}\label{Lpixel}} $\P$ be a digital picture and $p$ a corresponding $i$ or $o$-pixel, such that $p$ belongs to the exterior of $\P$. If $\P$ has two pixels that are D-connected to each other, and also 4-connected to $p$, then we say that $p$ is an {\it L-pixel} of $\P$.
\\

In Definition \ref{Lpixel}, if we write the pixel coordinates as (line, column) in Figure \ref{f6}, then $\P$ has exactly the following L-pixels: $(2,5)$, $(4,5)$, $(7,2)$, $(10,4)$, $(10,8)$, $(9,20)$ $(6,22)$, $(4,23)$, $(3,20)$, $(3,15)$, $(5,10)$, $(3,9)$ and $(2,8)$.   

{\Defi{\rm Given}\label{Legocurve}} a digital picture $\P$ with exterior $\E$, let $\L$ be the set of L-pixels of $\P$ that are 4-connected with $\E$. The {\it Lego curve} of $\P$ is a digital curve of minimal length that satisfies {\bf C2}, and has an exterior $\E'\subset\E$ such that $\E\setminus\E'\equiv\L$.
\\

We conclude this section by remarking that Definition \ref{Legocurve} {\it does not} require the Lego curve to be locally thin, not even free of self-intersections. However, they can only occur in the special form of {\it overlapping}, not {\it self-crossing}. The FUA does work in the former case, but can fail in the latter.

\section{Existence and uniqueness of the Lego curve}
\label{mainthm}

In this section we present our main theorem, which translates to the open source code {\tt lego.m}. We have written our proof like a {\it pseudocode}. 

{\thm Any digital picture $\P$ admits a unique corresponding Lego curve $L$.\label{Legothm}}
\ \\
\ \\
\underline{Proof:} We are going to construct a list $L:[0\dots n]\to\Z^2$. Take $\P$ as in Definition \ref{exterior}. Its pixels will be called {\it black}. However, we shall say that a pixel is {\it white} exactly when it belongs to the exterior of $\P$. In what follows, both the words {\it increase/decrease} mean ``by one''.

Consider the $i$-$o$ pixels of $\P$ with least ordinate $y$. Among them, name $p$ the one with the least abscissa $x$. It has coordinates $(p_y,p_x)$, $1<p_y<N$ and $1<p_x<M$. We start with $n=0$ and either $L(0)=p$ or $L(0)=p+(0,1)$, depending on whether $p$ is an L-pixel or not, respectively. At any step, the construction interrupts when $L(n)$ coincides with $L(0)$.
\ \\
\ \\
\underline{Step I}. Increase $n$ and define $L(n)$ as follows: 
\begin{description}
\itemsep = 0.0 pc
\parsep  = 0.0 pc
\parskip = 0.0 pc
\item{}(a) if at the southwest of $L(n-1)$ we have a black or an 

\ \ \ \ \ \ \ \ L-pixel, then it will be $L(n+1)$, and $L(n)$ will be the 

\ \ \ \ \ \ \ \ southern pixel;
\item{}(b) otherwise, if the south is black or an L-pixel then it will 

\ \ \ \ \ \ \ \ be $L(n)$; 
\item{}(c) if neither (a) nor (b) then take $L(n)=L(n-1)+(0,1)$.
\end{description}
\noindent
\underline{Step II}. Now $L(n)$ has just been defined. 
\\

If $L(n+1)$ is already defined, interchange the roles $x\leftrightarrow y$ by rotating the canvas counterclockwise. Increase $n$, redefine $p$ as $L(n)$ and repeat Steps I-II.

If $L(n+1)$ is not defined yet, then we look at $L(n)_y$. In the case it is $L(n-1)_y+1$, redefine $p$ as $L(n)$ and repeat Steps I-II. In the case it is $L(n-1)_y$, we can have 
\begin{description}
\itemsep = 0.0 pc
\parsep  = 0.0 pc
\parskip = 0.0 pc
\item{}(a) $L(n)$ and $L(n)+(1,0)$ both white; this situation is predicted 

\ \ \ \ \ \ \ in Step III. 
\item{}(b) $L(n)+(1,0)$ is black or an L-pixel; take $L(n+1)=$ 

\ \ \ \ \ \ $\,\,L(n)+(1,0)$, increase $n$, redefine $p=L(n)$ and repeat 

\ \ \ \ \ \ \ Steps I-II.
\item{}(c) Otherwise redefine $p=L(n)$ and repeat Steps I-II.
\end{description}
\noindent
\underline{Step III}. In the present orientation you cannot increase $L(n)_y$ any more. Then interchange the roles $x\leftrightarrow y$ by rotating the canvas clockwise. Clear $L(n)$, decrease $n$ and repeat Steps I-II. Note: if this step happens three times consecutively, then $\P$ consists of a single pixel. This case can be excluded beforehand.
\ \\
\ \\
\underline{Remark}: If $\P$ does not consist of a single row of black pixels, then there will be a least value $n_1\le n$ such that $L(n_1)$ has a greater ordinate than $L(0)$ with respect to the initial orientation.
\\

\begin{figure}[ht!]
\centerline{
\includegraphics[scale=0.22]{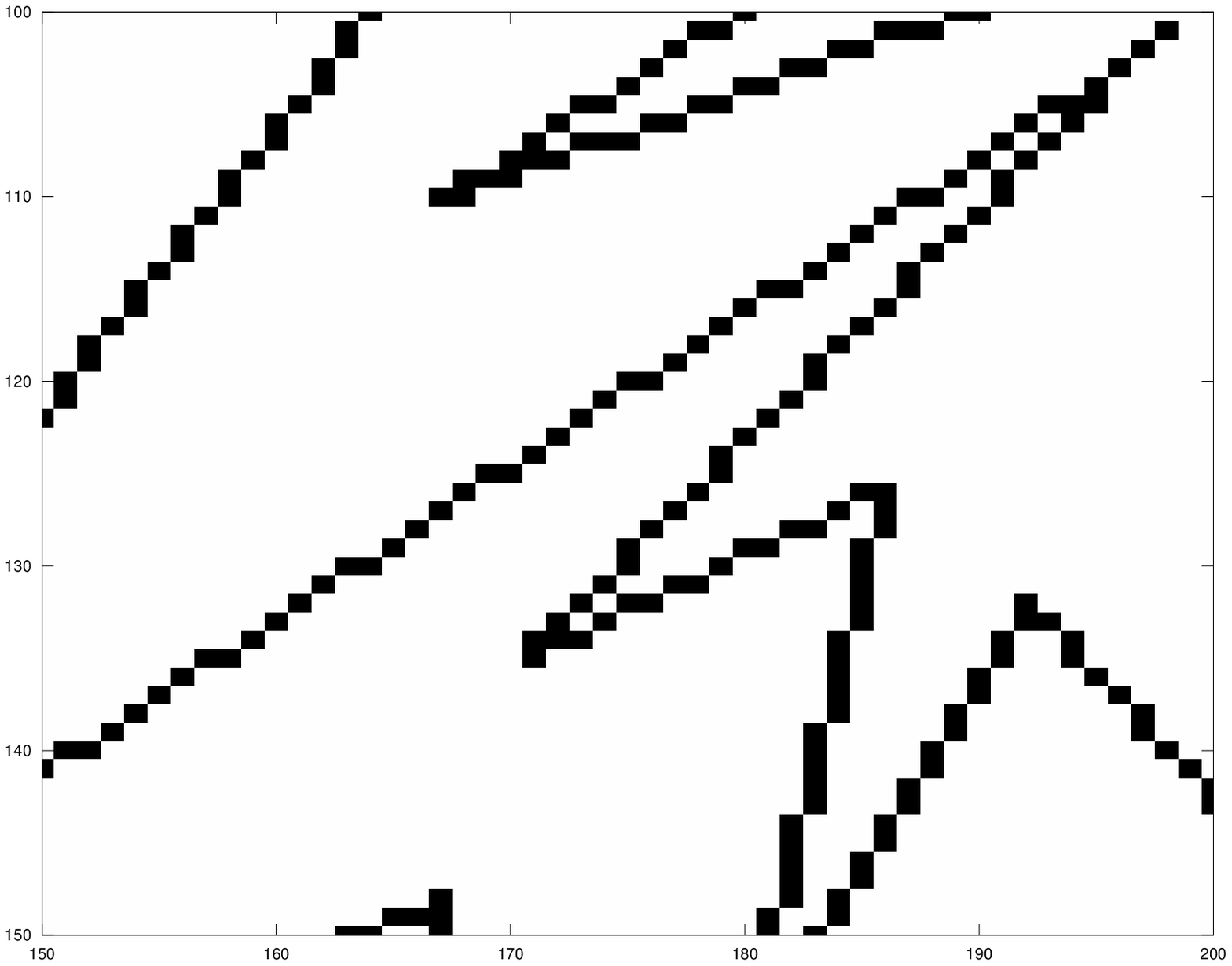}
\includegraphics[scale=0.22]{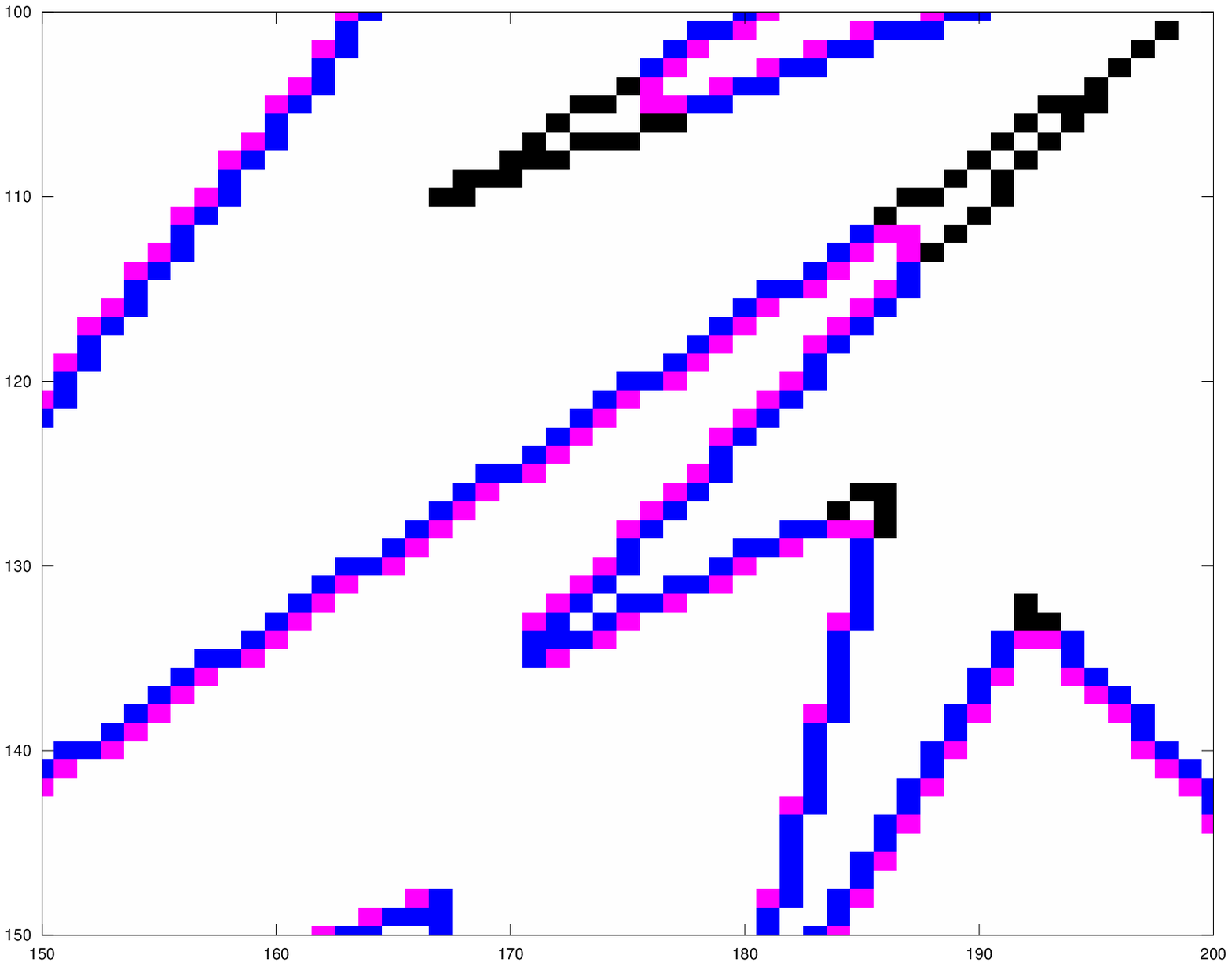}}
\caption{False self-intersections interpreted by CoTRA.}
\label{f7}
\end{figure}

The above construction is in fact what we called CoTRA. But the reader will notice a gap in it, as illustrated in Figure \ref{f7}. It is easily identified by ``trapped'' 4-connected L-pixels. Compare it with Figure \ref{f4}. But we easily go round this gap by identifying non-marked L-pixels and locally repeating the above construction.

However, our list $L$ may still have spikes. In this case, we recall Section \ref{reductions} and see that {\bf C2} trivially holds for $L$. Regarding {\bf C1}, it easily follows from {\bf C2} and {\bf C3}. In order to {\it get} {\bf C3}, we have to adapt the above construction as follows: if the roles $x\leftrightarrow y$ are interchanged consecutively twice, then lop off some newly defined $L(n)$ by checking if either $x$- or $y$-coordinates remained constant.

Of course, there remain two important questions that were not answered yet. The first is: why does the construction ever finish? In other words, why does $L(n)$ coincide with $L(0)$ for a certain positive $n$?

Notice that any two 8-connected pixels of $\P$, that are both connected to $\E$, can either project onto $Ox$ or $Oy$ injectively. We start with $Oy$ and keep on using it until injectivity fails. Then we change to $Ox$ and do the same to this new axis. This is exactly when the roles $x\leftrightarrow y$ are interchanged in the above construction. It begins with a local graph $x=x(y)$ so that pixels right {\it under} the graph belong to the exterior $\E$ of $\P$. As a matter of fact, this reasoning holds for the {\it whole} construction. Thus no pixel of $\E$, that is right under a local graph, is left out. Hence $L(n)$ will eventually coincide with $L(0)$ for a certain $n$. 

By the way, the same reasoning explains why $L$ contains the set $\L$ from Definition \ref{Legocurve}. This partially answers the second question: why $L$ fits Definition \ref{Legocurve}? It remains to check that it has minimal length among any other curve $L'$ with exterior $\E''\subset\E$ and $\E\setminus\E''=\L$. 

We know $\E=\E'\dot\cup\L=\E''\dot\cup\L$, thus $\E'=\E''$. Suppose there exists $p\in\L\setminus\L'$. Consequently, $p\in\P$ and is 4-connected with $\E$. Indeed, it cannot be only D-connected because then an L-pixel would have replaced it in the above construction. Either $p$ is 4-connected with $\E'=\E''$ or $p$ is simultaneously 4-connected with two different pixels $q,r\in\L$. This latter is the only possibility for $p\not\in\L'$, since $p\not\in\L$ and both $L,L'$ have to satisfy {\bf C2}. But then $q,r\in\L'$ have to be connected by a longer path than through $p$. Hence $L'$ cannot have minimal length.\hfill q.e.d.
\\

We finish this section by stating and proving a corollary, which is a weaker version of the {\it discrete Jordan curve theorem}. This is because we use an additional hypothesis $H$, described below. In \ref{ap2} we discuss some other theoretical results that can be derived from ours.

{\crl Let $\C:[0\dots n]\to S$ be a simple closed curve and denote its image by $\P$. Suppose $S$ and $\P$ are such that Definition \ref{exterior} applies. Additionally, the following holds:
\\
$H$: Any $p\in\P$ is connected to exactly two other elements of $\P$.
\noindent
Hence $S\setminus\P=\E\dot{\cup}\J$, where $\E$ is the exterior of both $\P$ and $\C$, and $\J$ the interior of both $\P$ and $\C$. Moreover, $\E$ is connected, and the same holds for $\J$.\label{dJct}}
\ \\
\ \\
\underline{Proof:} Since any $p\in\P$ is connected to exactly two other elements of $\P$, then $\C$ is thin and spike-free. The same will happen to $L$, constructed from $\P$ by means of Theorem \ref{Legothm}. Moreover, $L$ can only be tracked horizontally and vertically. Hence, it is not difficult to conclude that its image $\P'$ has the following property: $S\setminus\P'=\E'\dot{\cup}\J'$, where $\E'$ is connected, and the same holds for $\J'$. But $\E=\E'\cup\L$, where $\L$ is given by Definition \ref{Legocurve}. Moreover, $\E$ is connected. Now, $\J=S\setminus(\P\cup\E)=S\setminus(\P'\cup\E')=\J'$. Therefore, $\J$ is also connected.\hfill q.e.d.
\\

Notice that neither lists of Figure \ref{f1} fulfil the hypothesis of Corollary \ref{dJct}. 

\section{Results}
\label{ap1}

This present section is like a manual to the programme. Download {\tt legoloci.zip} from 
\\

{\tt https://sites.google.com/site/aefabris/codes}
\\
\\
and extract it in a folder. You will get {\tt loci.p}, {\tt lego.p} and some test-files. In order to draw our figures, we have chosen Xfig for Linux Ubuntu 12.04. They all have extension FIG, and the reader can either change them or even create new tests. However, our programmes {\it do not} read FIG-files. Whenever you create a figure, please export it in a format like TIF or JPG. 

We shall make the programmes {\tt loci.m} and {\tt lego.m} available in future. For the time being, users can run the p-codes in Matlab. To test the robustness of our algorithms, one can either edit the presented test files or even create new ones.

The source codes run both in Matlab and Octave. Table \ref{times} shows the average performance of these programmes in each software, providing you use 4GB of RAM, microprocessor Intel Core i5 3.2GHz, and operational system Linux Ubuntu 12.04. Table \ref{times} refers to Matlab 7.8.0 (R2009a) and Octave 3.2.4 (R2009). All test files have extension JPG, {\bf M} stands for Matlab and {\bf O} for Octave.  

\begin{table}[t]
\centering
\caption{Comparison to Computational Times in Seconds}{
\begin{tabular}{@{}lcccc@{}}\hline
&\multicolumn{1}{c}{\bf loci-M}
&\multicolumn{1}{c}{\bf lego-M}
&\multicolumn{1}{c}{\bf loci-O}
&\multicolumn{1}{c}{\bf lego-O}
\\ \hline
test1 & 0.30 & 0.80 & 0.34 & 2.18 \\
test2 & 0.36 & 1.17 & 0.51 & 4.03 \\
test3 & 0.29 & 0.63 & 0.34 & 2.91 \\
test4 & 0.29 & 0.76 & 0.39 & 3.46 \\
test5 & 0.56 & 2.65 & 1.25 & 8.40 \\
test6 & 0.30 & 0.89 & 0.41 & 2.32 \\
test7 & 0.36 & 0.59 & 0.43 & 7.51 \\
test8 & 0.30 & 0.93 & 0.39 & 3.86 \\
\hline
\end{tabular}}
\label{times}
\end{table}

\begin{table*}[ht!]
\tabcolsep22pt
\centering
\caption{Comparison to Seed-fill, Scan-conversion and Our approach}{
\begin{tabular}{@{}lccc@{}}\hline
&\multicolumn{1}{c}{\bf Seed-fill}
&\multicolumn{1}{c}{\bf Scan Conversion}
&\multicolumn{1}{c}{\bf Our approach} 
\\ \hline
Input data             & scanned/math defined         & math defined   & scanned/math defined \\
Implementation         & na{\" i}ve recursive: simple & complex        & simple               \\
Modelling space        & screen space                 & object space   & screen space         \\
Boundary modelling     & points/polygons              & polygons       & points               \\
Algorithm              & integer/floating point       & floating point & integer              \\
Applications areas     & image processing         & computer graphics  & computer graphics and\\
                       & CG painting systems      &                    & image processing     \\
Additional memory      & very large stack         & small/large stack  & no/small stack       \\
Hard/software design   & unsuitable hardware    &suitable hard/software&suitable hard/software\\
Device dependency      & requires GetPixelValue   & device independent & device independent   \\
Theoretical background & simple                   & simple             & relatively complex   \\
Robustness             & to be more robust        & most algorithms fail& robust              \\
Efficiency             & relatively slow          & most are fast      & relatively fast      \\
\hline
\end{tabular}}
\label{comparisons}
\end{table*}

Type {\tt loci} at the Matlab prompt. You will be asked to give a filename, for instance {\tt test3.tif}. Please write the full name {\it with} extension. After pressing the Enter-key, {\tt loci} will display the figure and also its inside according to FUA. The elapsed computational time is printed in the Matlab terminal window. It counts all commands that come right after {\it having stored} the image in the variable {\it img} (see Algorithm \ref{alg:FUA}), and right after {\it having displayed} its interior on the screen.  

Now the user gets the following message:
\\

{\sf Try CoTRA? Yes = 1; No = any key}
\\

By choosing 1, CoTRA is called to create the Lego curve as in the proof of Theorem \ref{Legothm}. A new figure will be displayed, but now its {\it inside} matches Definition \ref{interior}. Once again, the elapsed computational time is printed in the Matlab terminal window. It counts all commands the same way as done in the previous case.

Now we illustrate some outputs.   

\begin{figure}[ht!]
\centerline{
\includegraphics[scale=0.45]{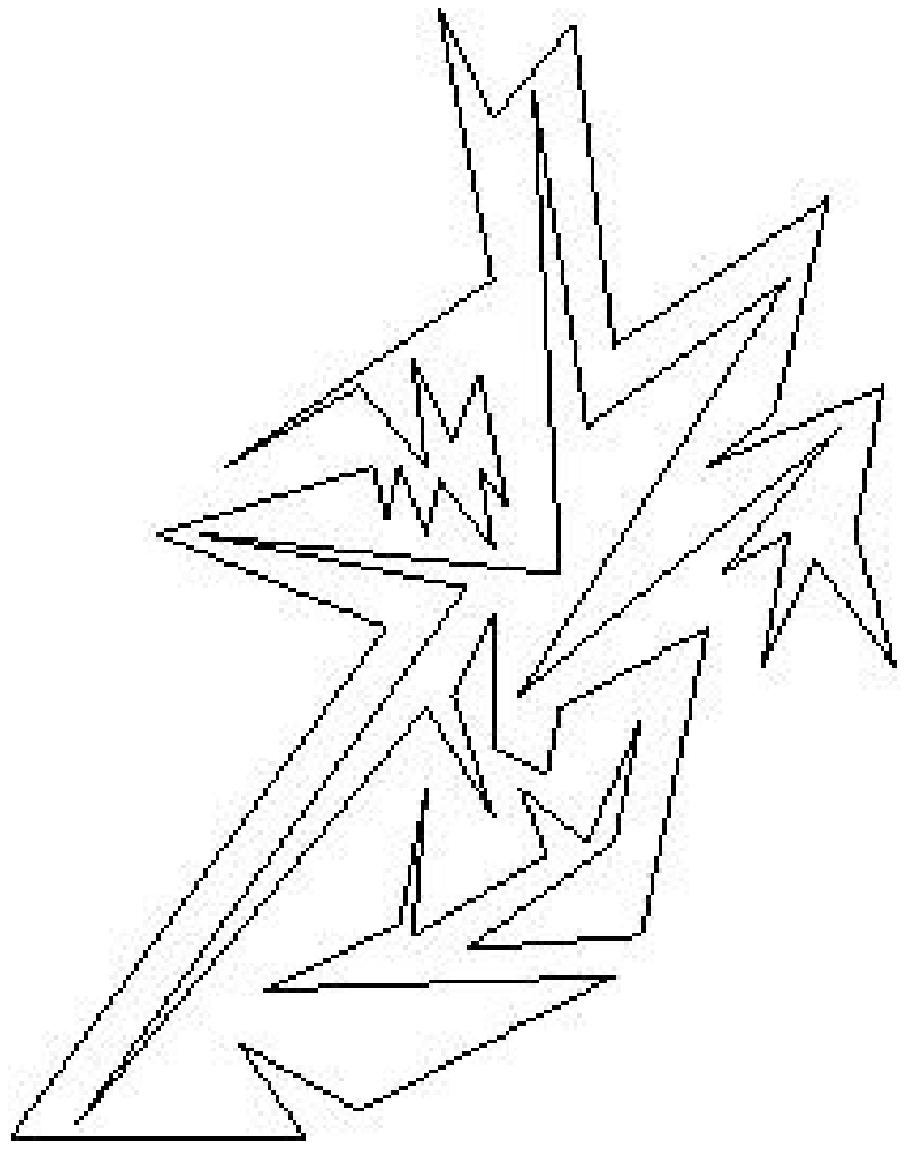}
\includegraphics[scale=0.70]{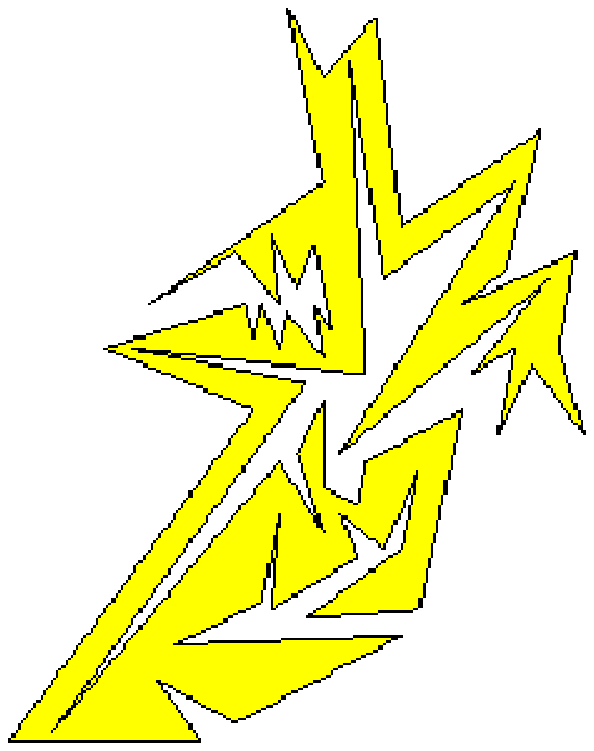}}
\caption{Filling up a challenging picture with CoTRA.}
\label{f8}
\end{figure}

At the Introduction, we mentioned that the marking phase of scan conversion requires careful attention to both geometric and numerical details. Figure \ref{f8} left was obtained by drawing a polygonal {\tt test4.fig} in Xfig. Like many other drawing softwares in computer graphics, the continuous mathematical structure is then discretized as a digital picture.

A detail of Figure \ref{f8} is zoomed in Figure \ref{f4}, which shows a self-intersection that occurs when we move from the {\it ideal simple (i.e. non-self-intersecting) continuous polygon} to the {\it digital picture}: the digital picture {\tt test4.jpg} is obtained by exporting {\tt test4.fig} to the JPG-format. As commented in Figure \ref{f4}, CoTRA finds a self-intersection at $(y,x)=(130,70)$. So {\tt test4.jpg} is filled in accordance with Definition \ref{interior}. 

\begin{figure}[ht!]
\centerline{
\includegraphics[scale=0.40]{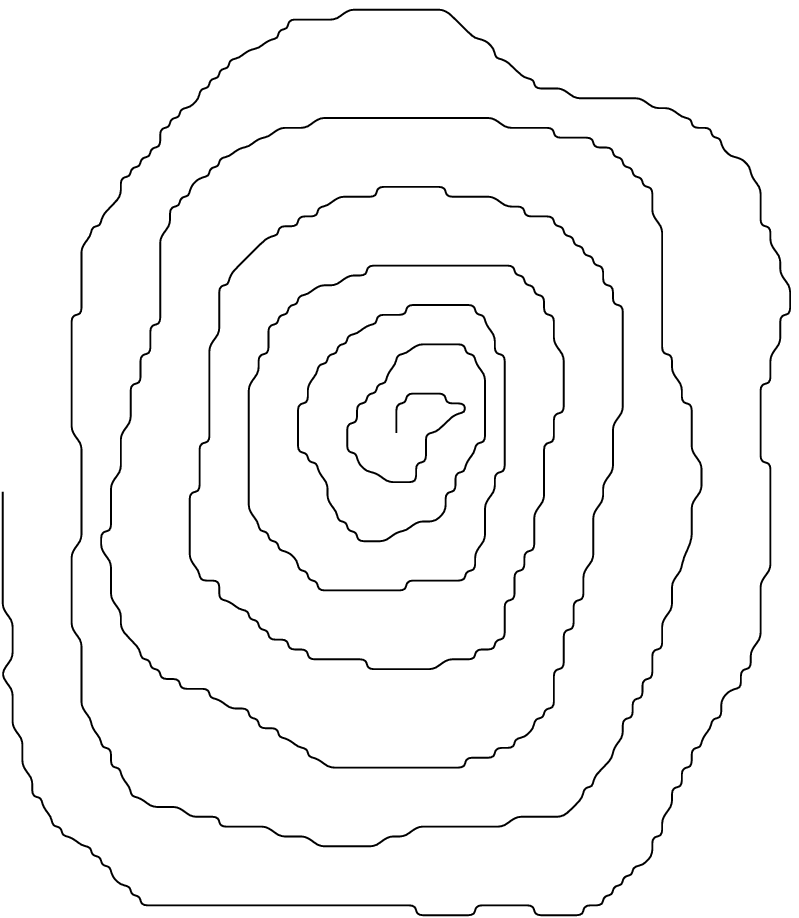}
\includegraphics[scale=0.40]{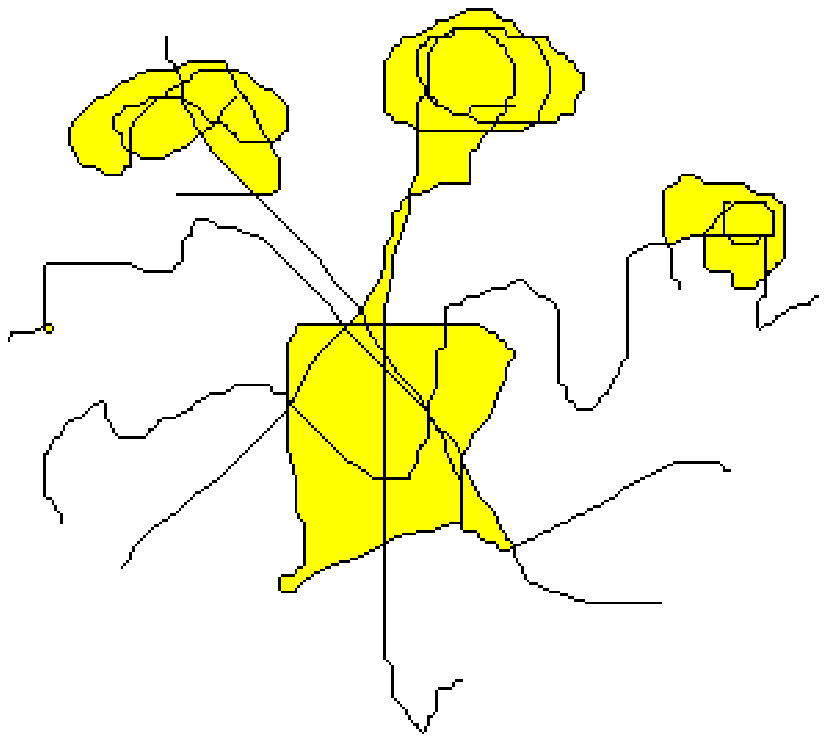}}
\caption{Examples of degenerate regions: empty and non-connected interiors.}
\label{f9}
\end{figure}

Figure \ref{f9} shows that parts with empty or non-connected interiors are correctly detected.

\section{Conclusions}
\label{concl}

In this paper, the difficulties in discretisation are acknowledged and tackled by casting the problem to be solved as a discrete integer problem from the very outset. 
Integer spaces are more amenable to analysis and proof. There the commonly employed geometric operations are translated into discrete versions, which allow a better control of robustness.

This approach enables the development of an integer-only algorithm, of which the main characteristics are: simple structure, no special cases to forecast, no extra memory allocation, and proved robustness. Namely, it can handle arbitrarily complex inputs. The source codes {\tt loci.m} and {\tt lego.m} were implemented without either recursive calls or repeat/while-loops. Only for-loops were allowed in the programmes. 
These characteristics can be forwarded to hardware implementation.

In fact, the main purpose of our work was to make faultless and fast algorithms that could also be implemented in hardware.

Regarding the efficiency of our present implementations, FUA alone has linear complexity with respect to $N\cdot M$. Our observations regarding CoTRA on numerous test scenes suggest quasi-linearity, but before any formalisation we need to check for improvements. Indeed, although CoTRA is relatively fast, we could improve its speed by exploring coherence tests in several ways. 

Moreover, FUA and CoTRA require input data that can either be scanned or constructed by a mathematical model. Therefore, these algorithms can be interactively used for applications in both Image Processing and Computer Graphics.

See Table \ref{comparisons} for a summary of seed-fill, scan-conversion and our approach.

\appendix

\section{Some Theory After This Work}
\label{ap2}

Any seed-fill algorithm easily finds the exterior $\E$ of a picture $\P\subset S$ directly from Definition \ref{exterior}. Hence we get the interior as $S\setminus(\P\cup\E)$. But the computational cost is huge because seed-fill approaches are recursive.

Theorem \ref{Legothm} constructs the Lego curve, which after some pruning is then applied to FUA (see Section \ref{mainthm}). 

However, the Lego curve is much more than just a way to get correct digital answers. It can help prove theorems in the {\it continuous} Euclidean plane. For instance, consider this strong version of Jordan's theorem, called {\it Sch\"oenflies Theorem}:

{\thm Let $S^1$ be the unitary circumference in $\R^2$ centred at the origin. Let $f:S^1\to\C\subset\R^2$ be a homeomorphism between $S^1$ and $\C$. Hence, there exists a homeomorphism $F:\R^2\to\R^2$ such that $F|_{S^1}=f$.\label{Schfthm}}
\ \\

An immediate consequence of Theorem \ref{Schfthm} is that the {\it interior} of $\C$, namely {\it the bounded component of} $\R^2\setminus\C$, is simply connected. It is a powerful result that, however, does not seem to admit a proof accessible to graduate students. But here we outline some ideas of which details could be carried out in future works. 

As depicted in Figure \ref{f6}, we could get a {\it continuous} Lego curve out of any grid in which the original $\C$ is covered by black pixels. Ideally, resolution becomes infinite when the pixel dimension $(1/n)\times(1/n)$ goes to zero, where $n$ is a positive integer. For each $n$, we get the continuous Lego curve $L_n$ that corresponds to $\C$. It is not difficult to describe $f_n:S^1\to L_n$ such that $f_n\to f$, because $L_n$ is tracked only horizontally and vertically. This same reason allows us to construct a homeomorphism $F_n:\R^2\to\R^2$, with $F_n|_{S^1}=f_n$. Finally, a limit process will then result in the general proof.

\bibliographystyle{model3-num-names}
\bibliography{fv}
\end{document}